# Tactics and Tallies: A Study of the 2016 U.S. Presidential Campaign Using Twitter 'Likes'

**Yu Wang**
Political Science
University of Rochester
Rochester, NY, 14627
ywang176@ur.rochester.edu

**Xiyang Zhang**
Psychology
Beijing Normal University
Beijing, 100875
zxy2013@mail.bnu.edu

**Jiebo Luo**
Computer Science
University of Rochester
Rochester, NY, 14627
jluo@cs.rochester.edu

## Abstract

We propose a framework to measure, evaluate, and rank campaign effectiveness in the ongoing 2016 U.S. presidential election. Using Twitter data collected from Sept. 2015 to Jan. 2016, we first uncover the tweeting tactics of the candidates and second, using negative binomial regression and exploiting the variations in 'likes,' we evaluate the effectiveness of these tactics. Thirdly, we rank the candidates' campaign tactics by calculating the conditional expectation of their generated 'likes.'

We show that while Ted Cruz and Marco Rubio put much weight on President Obama, this tactic is not being well received by their supporters. We demonstrate that Hillary Clinton's tactic of linking herself to President Obama resonates well with her supporters but the same is not true for Bernie Sanders. In addition, we show that Donald Trump is a major topic for all the other candidates and that the women issue is equally emphasized in Sanders' campaign as in Clinton's.

Finally, we suggest two ways that politicians can use the feedback mechanism in social media to improve their campaign: (1) use feedback from social media to improve campaign tactics within social media; (2) prototype policies and test the public response from the social media.

## Introduction

Twitter is playing an important role in connecting the presidential candidates with voters (Sanders 2016). Between September 18, 2015 and January 23, 2016, Hillary Clinton posted 1316 tweets, Bernie Sanders 1698 tweets, Donald Trump 2533 tweets, Ted Cruz 1309 tweets, and Marco Rubio 908 tweets.[1] These tweets constitute a valuable data source because they are explicitly political in nature, they are many, and, importantly, they carry feedback information from the voters in the form of 'likes.'

In this paper, we solve two problems. We first study the tweeting tactics in the tweets: we analyze who are mentioned in these tweets and what issues are raised. We then use negative binomial regression to evaluate the effectiveness of these tactics by exploiting the variations in 'likes,' which we refer to as tallies. Our study focuses on the five leading candidates: Hillary Clinton (D), Bernie Sanders (D), Donald Trump (R), Ted Cruz (R), Marco Rubio (R).[2]

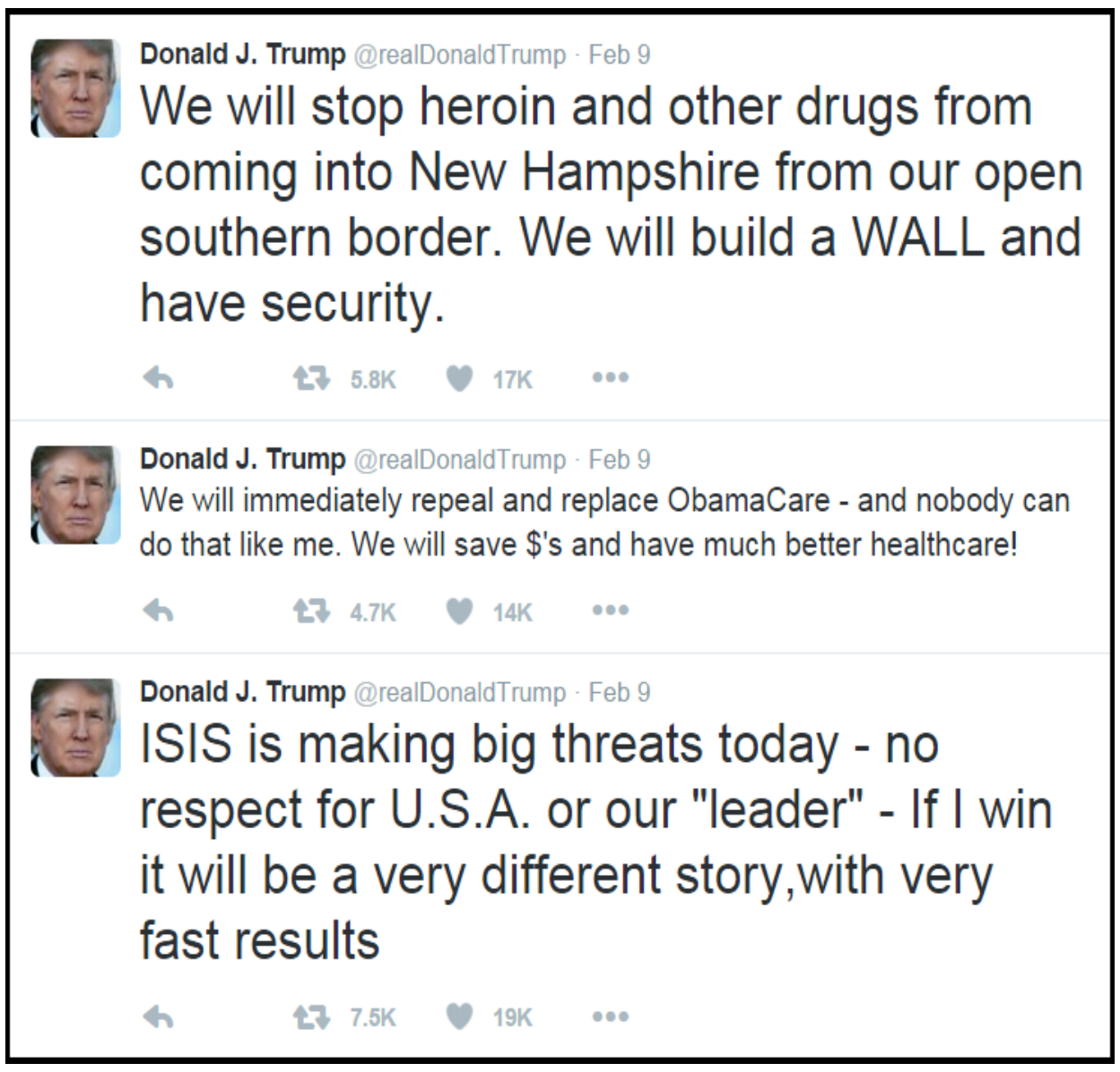

Figure 1: Selected tweets Donald Trump (R) posted on February 9th.

Figure 1 shall illustrate our points well. It shows three tweets that Donald Trump posted on February 9th, 2016, all of which are political in nature. The first tweet talks about drugs. The second tweet raises the issue of ObamaCare and points towards President Obama (D).[3] The third tweet is about the ISIS. Trump supporters responded to these three tweets differently, assigning to the third tweet the most 'likes' and to the second tweet the fewest 'likes.' By connecting topics with responses, we are therefore able to infer the effectiveness of the tactics.

Our approach has a distinct advantage over polls. Individ-uals surveyed in polls may not actually turn out to vote and even if they do vote they might change their mind and vote differently. By contrast, our data is built on strong revealed preference. Individuals voluntarily express their preferences, and thus the estimates we obtain are more reliable.

[1] We do not count retweets, as retweets do not have as a feature the number of 'likes.'

[2] This selection is based on both polling results and on candidates' performance in the Iowa caucus.

[3] Throughout, we follow the convention that Republican candidates are marked with (R) and Democratic candidates are marked with (D).

We organize our paper as follows. Section 2 presents related work. Section 3 introduces the dataset *US2016* and the methodology. Section 4 presents the tactics we uncover from the tweets. Section 5 evaluates these tactics. Section 6 ranks the campaign effectiveness of the five candidates. Section 7 concludes.

## Related Work

Our work builds upon previous research in electoral studies in political science and behavioral studies based on social media.

Political scientists have long studied the effects of campaigns and public debates. Many studies have found that campaign and news media messages can alter voters' behavior (Riker 1986; Iyengar and Kinder 1987). According to Gabriel S. Lenz, public debates help inform some of the voters about the parties' or candidates' positions on the important issues (Lenz 2009). In our work, we assume that tweets posted by the presidential candidates reveal their policy positions in various dimensions and that supporters reveal their policy preference by deciding whether or not to 'like' the tweets.

In a related strand, researchers have found that gender constitutes an important factor in voting behavior. One common observation is that women tend to vote for women, which political scientists refer to as gender affinity effect (King and Matland 2003; Dolan 2008). In our work, we test whether the women issue is emphasized in Hillary Clinton's campaign and evaluate its effectiveness.

There are also a large number of studies on using social media data to analyze and forecast election results. DiGrazia et al. (DiGrazia et al. 2013) find a statistically significant relationship between tweets and electoral outcomes. MacWilliams (MacWilliams 2015) suggests that a candidate's number of 'likes' in Facebook can be used for measuring a campaign's success in engaging the public. According to Williams and Gulati (Williams and Gulati 2008), the number of Facebook fans constitutes an indicator of candidate viability. According to O'Connor et al. (O'Connor et al. 2010), tweets with sentiment can potentially serve as votes and substitute traditional polling. Gayo-Avello, Metaxas and Mustafaraj (Gayo-Avello, Metaxas, and Mustafaraj 2011), on the other hand, report unsatisfactory performance of such predictions and advocate that scholarly research should be accompanied with a model explaining the predicative power of social media.

Substantively, our paper is closely related to two existing studies of the 2016 U.S. presidential election. Wang, Li and Luo (Wang, Li, and Luo 2016b) study the growth pattern of Donald Trump's followers. Wang, Li and Luo (Wang, Li, and Luo 2016a) use Twitter profile images to study and compare the demographics of the followers of Donald Trump and Hillary Clinton. The authors also study the effects of public debates on the number of Twitter followers. Our work uses both the number of candidate followers (as a control variable) and the number of 'likes' (as the dependent variable). Our contribution is to infer tactic effectiveness from these 'likes."

There are also quite a few studies modeling individual behaviors on social media. Lee et al. (Lee et al. 2015) model the decision to retweet, using Twitter user features such as agreeableness, number of tweets posted, and daily tweeting patterns. Mahmud, Chen, and Nichols (Mahmud, Chen, and Nichols 2013) model individuals' waiting time before replying to a tweet based on their previous replying patterns. Our study models the number of 'likes' that candidates' tweets receive. Our innovation is to use tweet-specific features instead of individual-specific features, as is done in the abovecited literature. A closely related work is Wang, et al. (Wang et al. 2016), which uses Latent Dirichlet Allocation (LDA) to extract tweet topics and models supporter preferences. In this paper, we reply on domain knowledge and search for specific topics using predefined keywords. As a result, we will have definitive labels.

## Data and Methodology

We use the dataset *US2016*, constructed by us with Twitter data. The dataset contains a tracking record of the number of followers for all the major candidates in the 2016 presidential race, including Hillary Clinton (D), Bernie Sanders (D), Donald Trump (R), Ted Cruz (R), and Marco Rubio (R). The dataset spans the entire period between September 18th, 2015 and January 23th, 2016, and covers four Democratic debates and four Republican debates.

**Dependent variable**

Our dataset *US2016* contains all the tweets that the five candidates posted during the same period and the number of 'likes' that each tweet has received. In Table 1, we report the summary statistics of the dependent variable: 'likes.'

Table 1: Summary statistics

| Variable | Mean | Std. Dev. | Min. | Max. | N |
|---|---|---|---|---|---|
| Hillary Clinton | 1702.454 | 1682.262 | 120 | 19935 | 1316 |
| Bernie Sanders | 2792.472 | 2748.807 | 204 | 44267 | 1698 |
| Donald Trump | 3757.479 | 2900.289 | 730 | 32652 | 2533 |
| Ted Cruz | 437.015 | 553.396 | 8 | 7991 | 1309 |
| Marco Rubio | 367.976 | 511.606 | 3 | 4856 | 908 |

To visualize these tallies, we plot the density distribution for each candidate, grouped by party affiliation. We align the x axis so that it is easy to compare the distribution both across candidates and across parties. We observe that in the Democratic party, Sanders' tweets tend to receive more 'likes' than Clinton's tweets. Among Republican candidates, Trump's tweets receive more 'likes' than Cruz and Rubio. Equally important, we observe large variations in the distribution for all the candidates.

**Explanatory Variables**

We believe part of the variations can be attributed to the topics embedded in the tweets: a more preferred topic generates more 'likes.' To operationalize this idea, we first multilabel each tweet for the following individual-based topics:

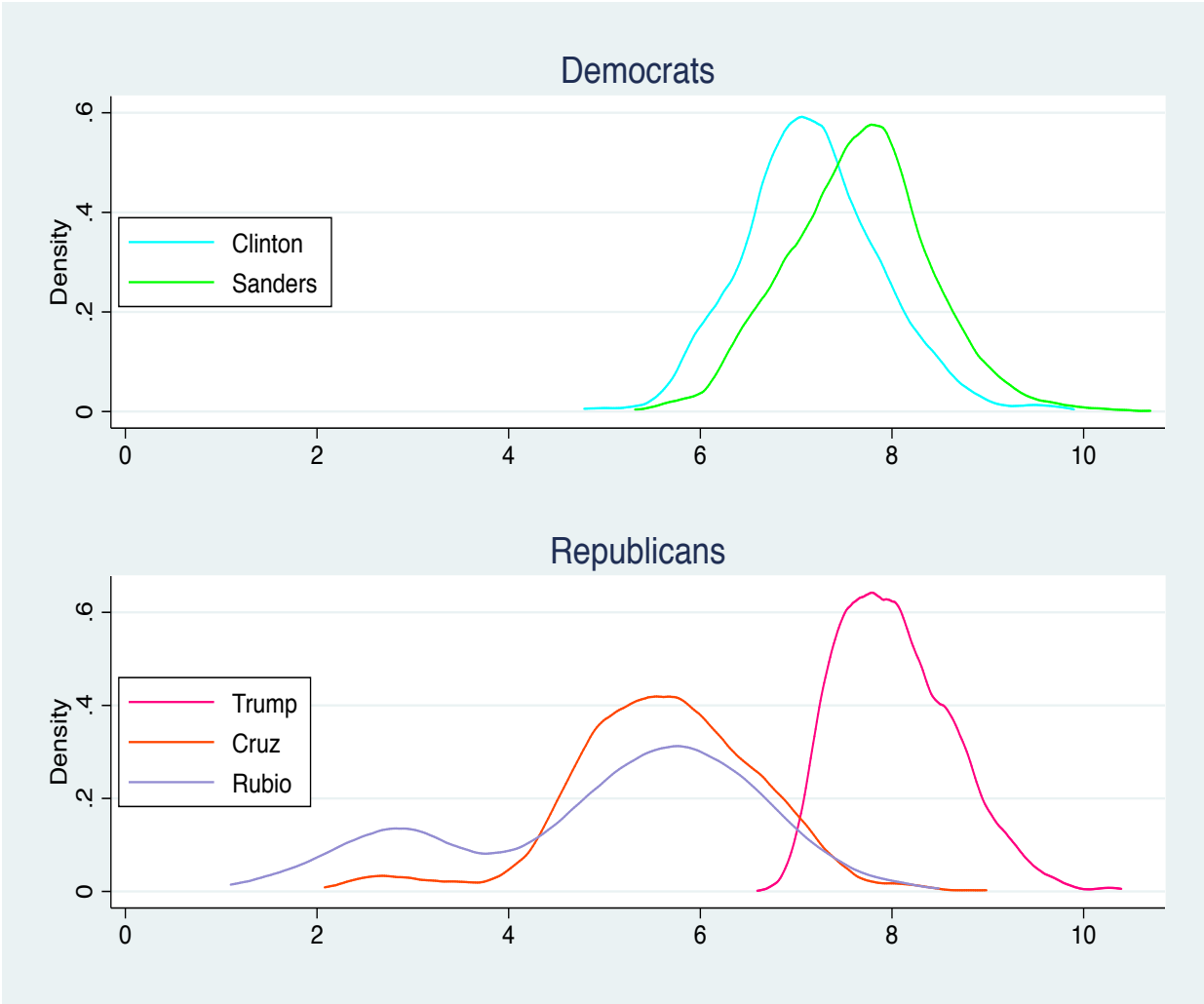

Figure 2: Distribution of 'Likes' in Log units.

President Obama (D), Hillary Clinton (D), Bernie Sanders (D), Martin O'Malley (D), Donald Trump (R), Ted Cruz (R), Marco Rubio (R), Jeb Bush (R), Ben Carson (R), Rand Paul (R), John Kasich (R), and Chris Christie (R).[4] We then multi-label each tweet for issue-based topics: ISIS, immigration, Iran, women, education, drugs, gun control, abortion, economy and the Wall Street.[5]

Topic features are binary. We derive these features using keyword matching. For example, we assign to the Obama topic 1 for tweets that contain "Obama" and assign 1 to the Rubio topic for tweets containing "marcorubio" (case-sensitive) or "Rubio." For issue topics, we first transform the tweets to lowercase before the matching procedure. For example, we assign 1 to the abortion topic for tweets that either contain "abortion" or "planned parenthood."

**Control Variables**

Following Wang, et al. (Wang et al. 2016), we control for the number of followers that each candidate has on Twitter. Intuitively, the more followers the candidate has the more 'likes' his or her tweets shall receive, as followers are more likely to notice the tweets posted by these candidates. On Figure 3, we report the growth of followers for the five candidates under study.

In addition to the number of followers, we also control for three tweet-specific features: the length of the tweet calculated as the number of words, whether the tweet contains a hyperlink, which is binary, and whether the candidate mentions himself or herself in the message.[6]

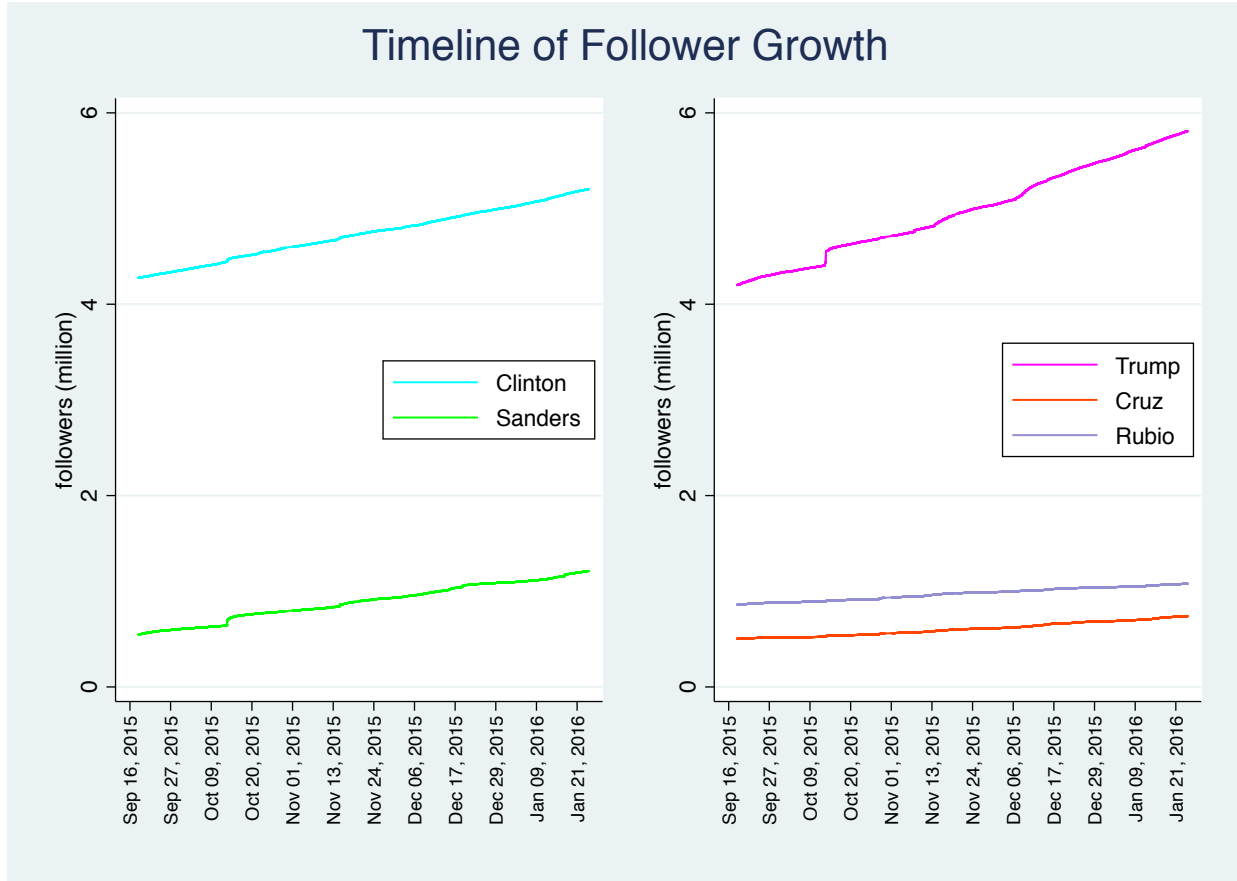

Figure 3: Growth of Candidates' Twitter Followers.

**Negative Binomial Regression**

Negative Binomial regression is the main workhorse in our study. Our dependent variable 'likes' is count data. In our MLE estimation, we formulate the link functions and the log-likelihood as follows (Greene 2008):

$$\begin{aligned}
\mu &= exp(\beta_0 + \beta_1 \text{Follower Count} + \beta_2 \text{Tweet Length} \\
&\quad + \beta_3 \text{Hyperlink} + \beta_4 \text{Self Referencing} \\
&\quad + \boldsymbol{\gamma} \cdot \textbf{Political Figures} + \theta \cdot \textbf{Political Issues} + \nu) \\
e^{\nu} &\sim Gamma(1/\alpha, \alpha) \\
p &= 1/(1+\alpha\mu) \\
m &= 1/\alpha \\
lnL &= \sum_{j=1}^{n} [ln(\Gamma(m+y_j) - ln(\Gamma(y_j+1)) - ln(\Gamma(m)) \\
&\quad + \text{m}ln(p_j) + y_j ln(1-p_j)]
\end{aligned}$$

where *Follower Count* is the number of Twitter followers (in millions) averaged over the day when the tweet is posted, *Tweet Length* is the number of words that the tweet contains. *Hyperlink* is binary and is 1 if the tweet contains an external link. **Political Figures** is a vector of 11 binary variables. The binary variable is 1 when the political figure is mentioned in the tweet and is 0 otherwise. **Political Issues** is a vector of 10 binary variables. The binary variable is 1 when the political issue is raised in the tweet and is 0 otherwise. Lastly, $\alpha$ is a measure of dispersion of the data.

**Forward-Stepwise Selection**

Out of the total 21 features, we select the top five topics using the Forward-Stepwise selection algorithm (Hastie, Tibshirani, and Friedman 2013). Such a selection allows us to uncover the most important topics and to order them accordingly.[7]

[4]By the end of the New Hampshire primary, Martin O'Malley (D), Rand Paul (R), and Chris Christie (R) have quit the race.

[5]The selection of political figures is based on the poll performance. For poll data, please refer to http://elections.huffingtonpost.com/pollster#2016-primaries. Our selection of political issues follows the Bing Political Index. Available at https://blogs.bing.com/search/2015/12/08/the-bing-2016-election-experience-how-do-the-candidates-measure-up/.

[6]Note that here we treat candidates' self-referencing behavior as a control variable rather than as a distinct topic.

[7]While we use loglikelihood as our selection criterion, using the Akaike information criterion (AIC) would yield exactly the same result, as AIC = 2k - 2loglike.

**Forward-Stepwise Selection (S, k)**

```
Add in control variables;
Initialize the topic set S = {topic_i, ∀i};
Initialize the selection set to S* = ∅;
for i from 1 to k: // k≤ |S|;
        topic= argmax logPr(Θ|y, topic ∈ S);
        S* = S* ∩ {topic};
        S = S\{topic};
end for;
```

### Tactic Evaluation

We propose the following metric to measure the candidates' tactics:

$$Eval(i) = \sum_j f(j)p(j|i, \sum I(j) > 0)$$

where i denotes candidate, j denotes topic, f(j) denotes the marginal effects of topic j that we derive using negative binomial regression, and p(j—i, $\sum I(j) > 0$) denotes the conditional probability of candidate i engaging in topic j given the candidate is raising at least one topic. We learn this conditional distribution from the tweets.

## Tactics

In this section, we report on the tactics that we uncover from the candidates' tweets. We first present the tactics with regard to political figures and then present tactics on policy issues. We follow the convention to color Democrats blue and Republicans red. Whenever possible, we draw references from the political science literature.

### Political Figures

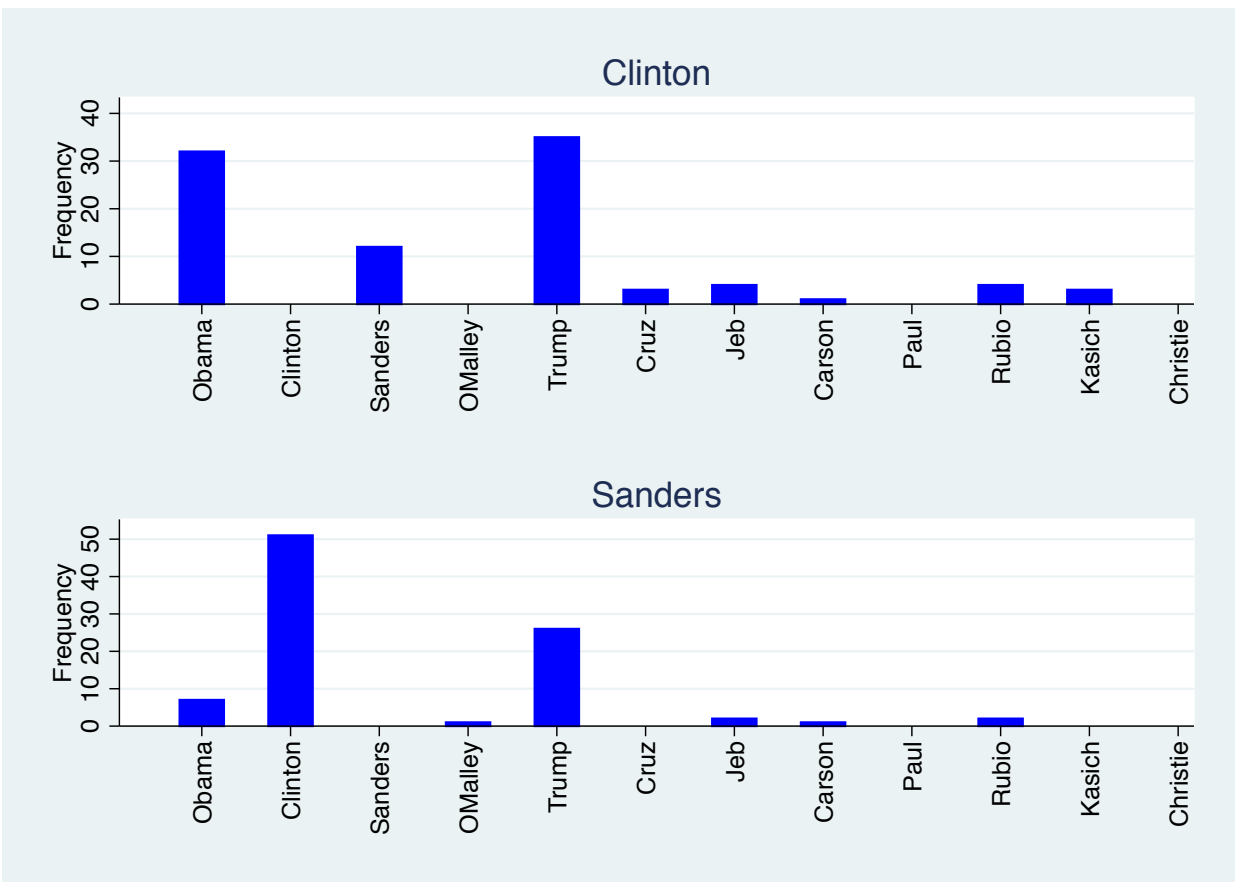

Figure 4: Distribution of Tweets on Political Figures for Hillary Clinton and Bernie Sanders.

In Figure 4, we present the histogram of the individual-based tweet topics. One immediate observation is that Hillary Clinton focuses on Donald Trump and President Obama, and that Bernie Sanders focuses on Hillary Clinton and Donald Trump, with President Obama being a distant third.[8] Other political figures barely receive attention.

In Figure 5, we present the histogram of the individual-based tweet topics for Trump, Cruz and Rubio. One sharp contrast among the three candidates is that Donald Trump virtually engages with all the political figures, whereas Ted Cruz and Marco Rubio focus their attacks on President Obama and Hillary Clinton.[9]

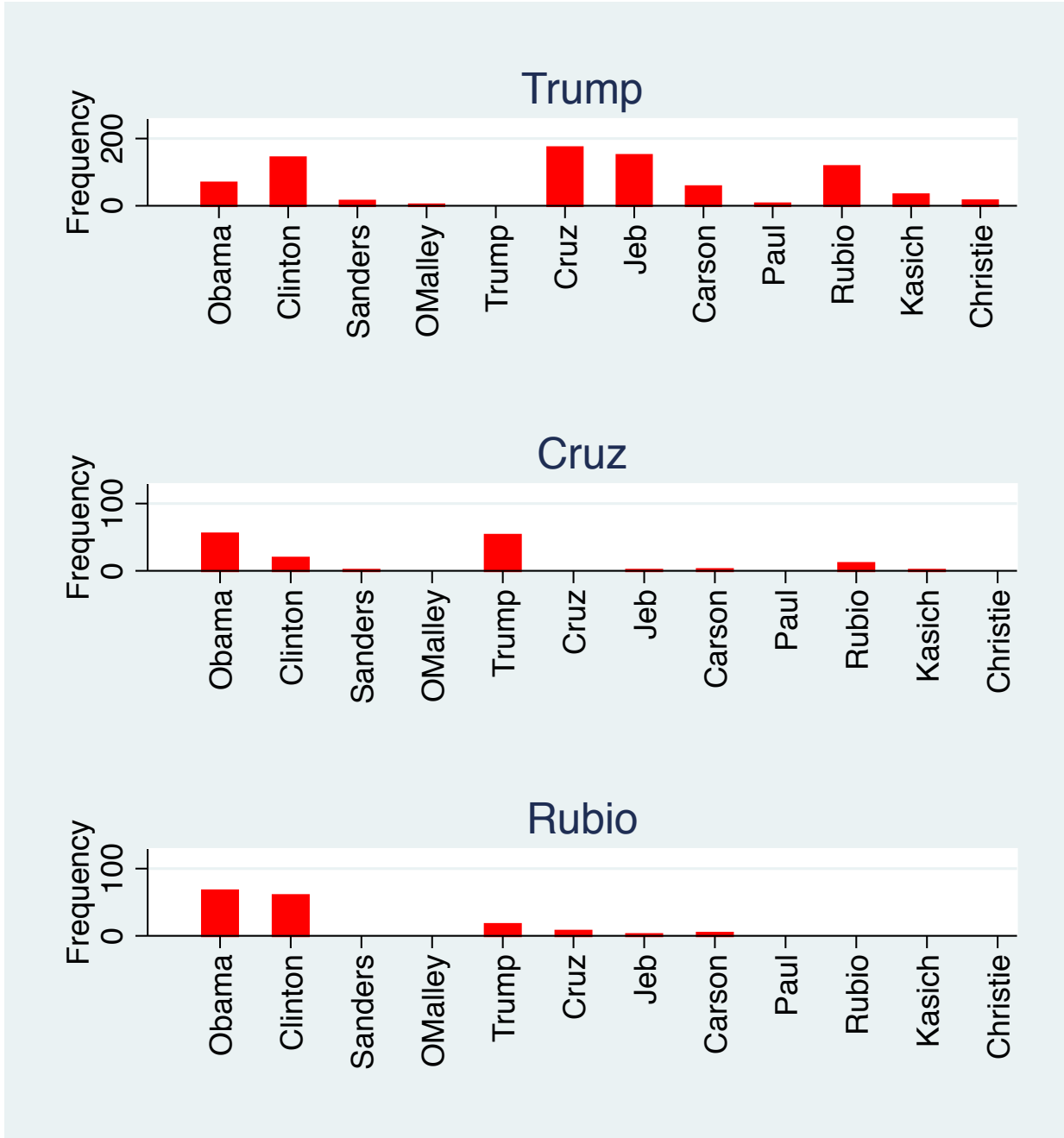

Figure 5: Distribution of Tweets on Political Figures for Donald Trump, Ted Cruz and Marco Rubio.

### Political Issues

In Figure 6, we report on the political issues that Democratic candidates focus on. Hillary Clinton focuses on the women issue and on gun control; Bernie Sanders focuses on the economy, the Wall Street, and education.

Putting women at the center of her campaign is one of the key strategies in Hillary Clinton's campaign.[10] Passing tougher gun control regulations is also high on her agenda, as Clinton tries to follow President Obama who has been pushing for gun controls. Bernie Sanders is less progressive on that issue. As *New York Times* writes, Sanders' mixed record on gun controls is one of his major vulnerabilities in the campaign.[11] Our data shows that Sanders' strategy is to avoid this issue.

[8]Note that while Clinton's focus is on both Trump and President Obama, she is attacking Trump while embracing President Obama. This is not immediately readable from the histogram.

[9]For a discussion of Marco Rubio's attacks on President Obama, please see. http://abcnews.go.com/Politics/marco-rubio-defends-repeated-attack-president-obama-republican/story?id=36760445.

[10]For a detailed historical discussion and a comparison with Clinton's bid for the White House in 2008, see http://thehill.com/homenews/campaign/219058-hillary-clinton-puts-womens-rights-at-center-of-her-agenda.

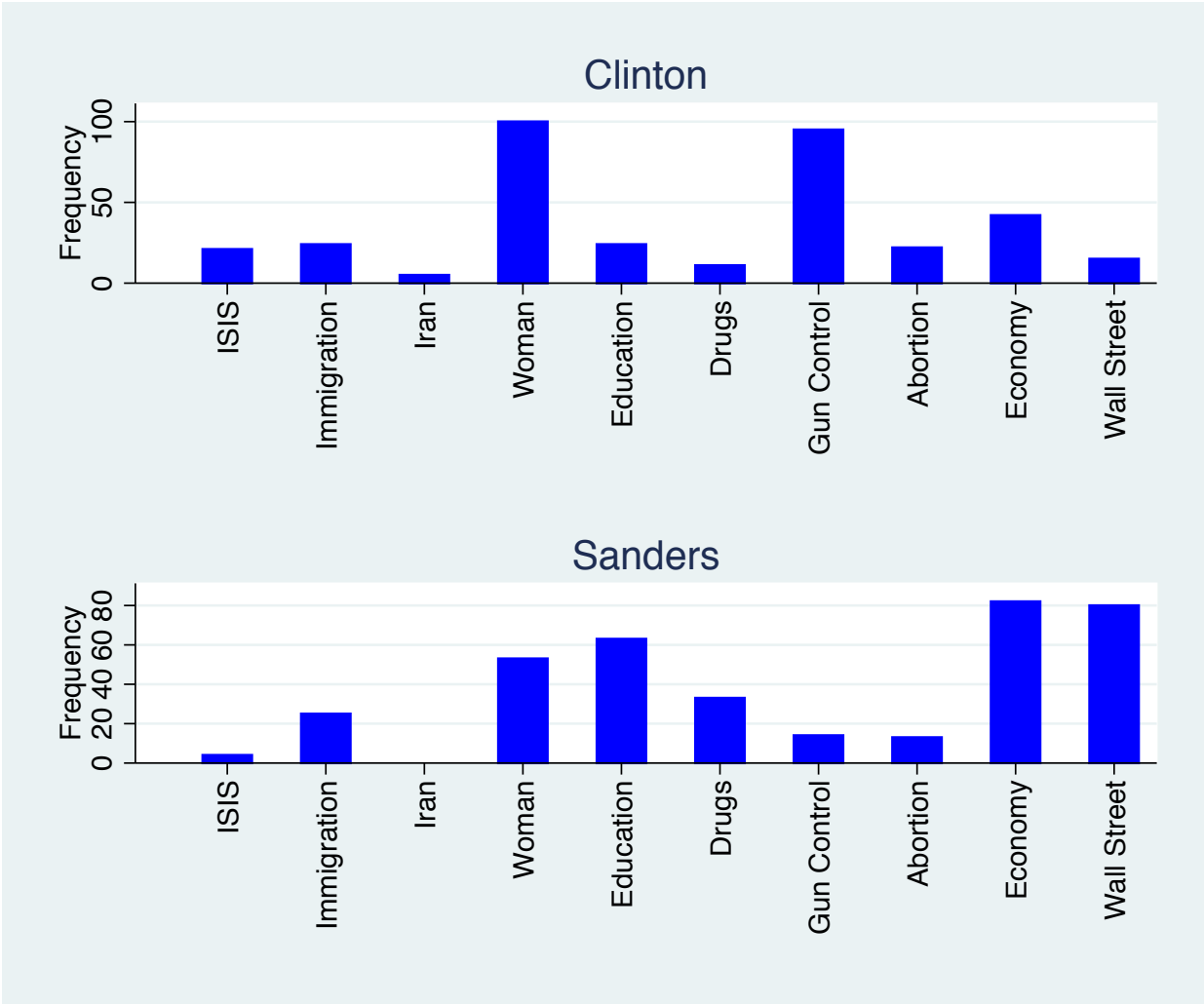

Figure 6: Distribution of Tweets on Political Issues for Hillary Clinton and Bernie Sanders.

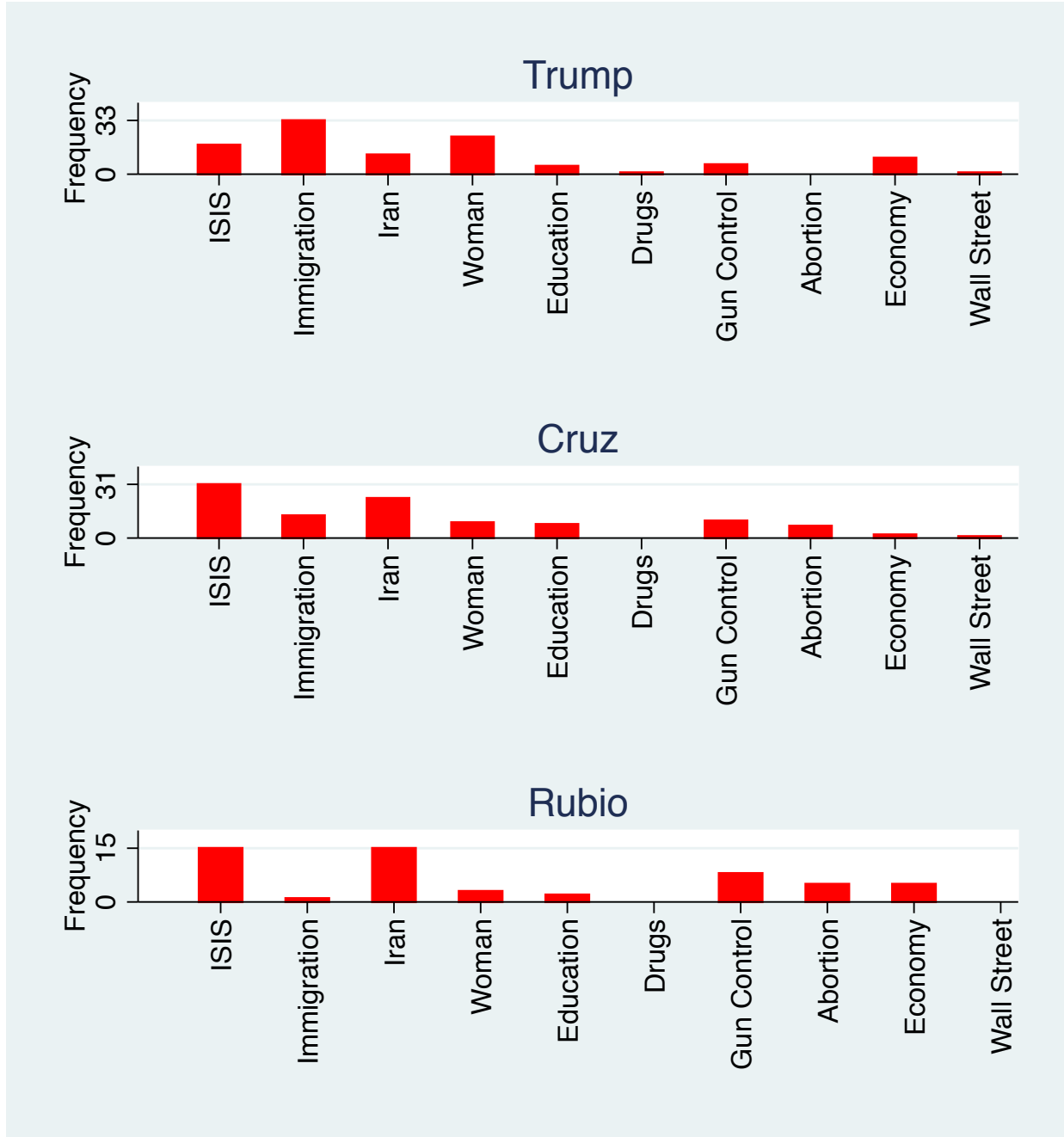

Figure 7: Distribution of Tweets on Political Issues for Donald Trump, Ted Cruz and Marco Rubio.

In Figure 7, we report on the political issues that Republican candidates focus on. Trump's most frequent topic is immigration, which is not surprising given Trump is the original proposer for constructing a wall between the United States and Mexico. For Ted Cruz and Marco Rubio, it is the ISIS. The overall pattern is that the Republican candidates focus on foreign affairs, and the Democrats focus on domestic affairs.

## Tallies

In this section, we first use negative binomial regression to connect these tactics to the tallies of 'likes' and measure which tactic fairs best in terms of generating 'likes.' Secondly, we use Forward-Stepwise selection to select and rank the top 5 salient topics for the candidates.

**Negative Binomial regression**

We report the estimated coefficients in Table 2. Each column represents one candidate and each row represents one topic or one control variable. Cells are missing when candidates do not discuss the corresponding topic (represented by rows).

We shall only interpret a few highlights here, but our results are much richer. First, we find the mention of President Obama increases the number of 'likes' for Clinton but not so for Sanders. This is strategically important when it comes to how candidates position themselves. While President Obama has not explicitly endorsed either Clinton or Sanders [12], voters tend to identify him more with the former Secretary of State than with Senator Sanders. Now in retrospect, Sanders' strategy of downplaying President Obama in his campaign is wise.

We also observe that focusing on attacking President Obama does not serve Cruz or Rubio well. This is an important discovery, because, unlike Sanders, both these candidates are investing heavily on the Obama topic. And we are showing this strategy is not generating many 'likes' among their supporters.

On the ISIS issue, our results show that all the Republican candidates are drawing more 'likes,' and that neither Democratic candidate wins on this issue. On immigration, Donald Trump (R) is clearly the winner. Hillary Clinton (D) and Bernie Sanders (D) both suffer on this topic. It is therefore no accident that the most frequent issue topic in Trump's campaign is immigration: it serves him well.

For Donald Trump (R), our results are consistent with (Wang et al. 2016), which finds that Trump supporters respond well to attacks on Democrats but less positively towards attacks on fellow Republicans, such as Jeb Bush. Our study shows that the result still holds at individual level and suggests that Trump should shift his attention away from fellow Republicans and focus more on Democrats.

For Rubio (R), one topic that resonates particularly well and sets himself apart from others is abortion. During the eighth Republican debate, Rubio attacks Clinton (D) and

[11]http://www.nytimes.com./politics/first-draft/2016/01/08/obamas-pledge-on-gun-control-leaves-bernie-sanders-with-something-to-prove/?_r=0.

[12]This is stated by the White House Chief of Staff Denis McDonough. Available at http://www.reuters.com/article/us-usa-election-obama-idUSKCN0UO0L720160110.

Table 2: Negative Binomial Regression

| | Clinton | Sanders | Trump | Cruz | Rubio |
|---|---|---|---|---|---|
| Likes | | | | | |
| Followers | 0.30*** | 0.43*** | 0.57*** | 3.88*** | 3.48*** |
| | (0.072) | (0.087) | (0.023) | (0.34) | (0.61) |
| Length | -0.015*** | -0.0062 | 0.015*** | 0.056*** | 0.071*** |
| | (0.0040) | (0.0037) | (0.0018) | (0.0040) | (0.0075) |
| Http | -0.27*** | -0.52*** | 0.11*** | 0.44*** | 0.45*** |
| | (0.045) | (0.036) | (0.025) | (0.061) | (0.083) |
| **Obama** | **0.33*** | **-0.16** | **0.39***** | **0.13** | **0.12** |
| | (0.15) | (0.27) | (0.070) | (0.13) | (0.15) |
| Clinton | -0.45*** | 0.28 | 0.28*** | 0.91*** | 0.37* |
| | (0.039) | (0.14) | (0.046) | (0.25) | (0.17) |
| Sanders | 0.084 | 0.67*** | 0.19 | | |
| | (0.34) | (0.14) | (0.14) | | |
| OMalley | | -0.19 | 0.39 | | |
| | | (0.66) | (0.28) | | |
| **Trump** | **0.43***** | **1.10***** | **-0.34***** | **1.12***** | **2.62*** |
| | (0.13) | (0.15) | (0.021) | (0.24) | (1.10) |
| Cruz | -0.17 | | -0.096 | -0.52*** | 1.89* |
| | (0.66) | | (0.063) | (0.051) | (0.78) |
| Carson | | 1.34* | -0.25** | 0.88 | 0.79 |
| | | (0.66) | (0.078) | (0.78) | (0.56) |
| Jeb | 0.33 | 0.075 | -0.11* | | -3.00* |
| | (0.38) | (0.46) | (0.047) | | (1.36) |
| Rubio | -0.30 | -0.73 | -0.12 | 0.67 | -1.00** |
| | (0.66) | (0.49) | (0.066) | (0.39) | (0.31) |
| Kasich | -0.52 | | -0.11 | | |
| | (0.38) | | (0.10) | | |
| Rand Paul | | | 0.028 | | |
| | | | (0.20) | | |
| Christie | | | -0.42* | | |
| | | | (0.16) | | |
| **ISIS** | **-0.011** | **0.20** | **0.48***** | **0.50***** | **1.18***** |
| | (0.15) | (0.33) | (0.12) | (0.14) | (0.29) |
| Immigration | -0.40** | -0.32* | 0.18* | -0.41 | -0.17 |
| | (0.14) | (0.13) | (0.090) | (0.22) | (1.10) |
| Iran | 0.15 | | 0.22 | 0.073 | 0.063 |
| | (0.32) | | (0.14) | (0.17) | (0.30) |
| **Women** | **0.24***** | **0.48***** | **-0.042** | **0.27** | **-0.23** |
| | (0.071) | (0.096) | (0.10) | (0.26) | (0.64) |
| Education | -0.14 | 0.30*** | 0.65* | -0.021 | 0.53 |
| | (0.14) | (0.085) | (0.29) | (0.28) | (0.78) |
| Drugs | 0.083 | 0.023 | -0.76 | | |
| | (0.21) | (0.12) | (0.48) | | |
| Gun Control | 0.12 | 0.62*** | 0.61** | 0.19 | 0.39 |
| | (0.073) | (0.18) | (0.20) | (0.25) | (0.40) |
| Abortion | 0.077 | 0.36 | | 0.13 | 1.30** |
| | (0.15) | (0.20) | | (0.30) | (0.49) |
| Economy | -0.27* | -0.23** | 0.041 | 0.11 | 1.00* |
| | (0.11) | (0.074) | (0.15) | (0.55) | (0.49) |
| Wall Street | -0.63*** | -0.19* | -0.64 | -1.72* | |
| | (0.18) | (0.077) | (0.48) | (0.79) | |
| Constant | 6.59*** | 7.79*** | 5.23*** | 2.54*** | 0.91 |
| | (0.35) | (0.11) | (0.12) | (0.22) | (0.59) |
| $\alpha$ | 0.44*** | 0.43*** | 0.23*** | 0.60*** | 1.20*** |
| | (0.016) | (0.014) | (0.006) | (0.022) | (0.050) |
| ln($\alpha$) | -0.83*** | -0.85*** | -1.48*** | -0.51*** | 0.19*** |
| | (0.037) | (0.032) | (0.028) | (0.037) | (0.041) |
| Observations | 1306 | 1688 | 2445 | 1252 | 889 |
| *AIC* | 21568.1 | 29468.6 | 42921.2 | 17215.4 | 11979.5 |

Standard errors in parentheses
$^{*}$ $p < 0.05$, $^{**}$ $p < 0.01$, $^{***}$ $p < 0.001$

Sanders (D) for not raising the issue during the Democratic debates.[13] This criticism is consistent with our data: both Clinton and Sanders have downplayed the topic of abortion. Rubio is also strategically right in emphasizing his pro-life stance, when he explains that he would rather lose an election than be wrong about abortion. Our study shows that the topic of abortion resonates well with Rubio's supporters.

To numerically evaluate the effectiveness of each topic for each candidate, we calculate the marginal effects of that topic, while holding all other variables to their mean. We then plot the estimates with 95% confidence intervals in Figures 8-12. They are reported at the end of the paper.

Using likelihood ratio test on $\alpha$, we are further able to confirm the existence of over-dispersion, and thus confirm that negative binomial regression is more appropriate than Poisson regression.

**Topic Selection**

We use the Forward-Stepwise selection algorithm to sequentially select the five most salient topics for each candidate. The selected topics are the ones that influence the number of 'likes' the most. We present the results in Tables 3 and 4.

Table 3: Topic Selection (D)

| | Topic | Clinton | Topic | Sanders |
|---|---|---|---|---|
| Likes | | | | |
| 1 | Trump | 0.42*** | Trump | 1.10*** |
| | | (0.13) | | (0.15) |
| 2 | Economy | -0.27* | Women | 0.54*** |
| | | (0.11) | | (0.092) |
| 3 | Wall Street | -0.65*** | Gun | 0.62*** |
| | | (0.17) | | (0.18) |
| 4 | Women | 0.23** | Education | 0.29*** |
| | | (0.070) | | (0.085) |
| 5 | Immigration | -0.39** | Economy | -0.25*** |
| | | (0.14) | | (0.075) |
| Constant | | 6.66*** | | 7.81*** |
| | | (0.34) | | (0.11) |
| $\alpha$ | | 0.44*** | | 0.43*** |
| | | (0.016) | | (0.014) |
| Constant | | -0.82*** | | -0.84*** |
| | | (0.037) | | (0.032) |
| N | | 1306 | | 1688 |
| *AIC* | | 21556.7 | | 29473.3 |

Standard errors in parentheses
$^{*}$ $p < 0.05$, $^{**}$ $p < 0.01$, $^{***}$ $p < 0.001$

Our study shows that Donald Trump (R) is on the top-5 list of all other candidates. It also shows that the ISIS topic is on the top lists of all the Republican candidates but is not prioritized either by Clinton or Sanders. *Women* is an important topic for Hillary Clinton (D). But it ranks even higher for Bernie Sanders (D). This is consistent with the exit poll results after the New Hampshire primary, in which Sanders won more women votes than Clinton and 69% of

[13] http://www.npr.org/sections/thetwo-way/2016/02/06/465880648/the-8th-republican-debate-in-100-words-and-3-videos.

women under 45 backed Sanders.[14]

Table 4: Topic Selection (R)

| | Topic | Trump | Topic | Cruz | Topic | Rubio |
|---|---|---|---|---|---|---|
| Likes | | | | | | |
| 1 | Obama | 0.39*** | Trump | 1.10*** | ISIS | 1.20*** |
| | | (0.070) | | (0.24) | | (0.29) |
| 2 | Clinton | 0.30*** | Clinton | 0.92*** | Trump | 2.56* |
| | | (0.045) | | (0.25) | | (1.11) |
| 3 | ISIS | 0.49*** | ISIS | 0.49*** | Abortion | 1.26* |
| | | (0.12) | | (0.14) | | (0.50) |
| 4 | Educt. | 0.79** | Rubio | 0.65 | Cruz | 1.50* |
| | | (0.28) | | (0.39) | | (0.64) |
| 5 | Gun | 0.56** | Immigt. | -0.42 | Economy | 0.97 |
| | | (0.20) | | (0.22) | | (0.50) |
| Const. | | 5.26*** | | 2.51*** | | 0.50 |
| | | (0.11) | | (0.21) | | (0.57) |
| $\alpha$ | | 0.23*** | | 0.60*** | | 1.22*** |
| | | (0.006) | | (0.022) | | (0.050) |
| Const. | | -1.46*** | | -0.51*** | | 0.20*** |
| | | (0.028) | | (0.037) | | (0.041) |
| N | | 2445 | | 1252 | | 889 |
| *AIC* | | 42936.5 | | 17205.2 | | 11974.0 |

Standard errors in parentheses
$^{*}\ p < 0.05$, $^{**}\ p < 0.01$, $^{***}\ p < 0.001$

## Tactic Evaluation

In this section, we evaluate the candidates' tactics by connecting their tweeting tactics and the 'likes' tally. We just observed that Donald Trump (R) is receiving many 'likes' for his comments on President Obama (D) and Hillary Clinton (D), but meanwhile he is also spending much time attacking Jeb Bush and Marco Rubio, which is not generating positive response. Senator Sanders (D), almost knowingly, avoids talking about President Obama to his benefits but his focus on the Wall Street is not generating positive response.

To solve this problem, we provide a unified approach to evaluate and rank the candidates' tactics, using the metric proposed in Section 3.

$$Eval(i) = \sum_j f(j)p(j|i, \sum I(j) > 0)$$

In Table 5, we report the evaluation results. Our calculation suggests that Sanders, who cunningly avoids talking about President Obama and focuses on Trump instead, has the best performing tactic. Hillary Clinton comes last. This can be attributed to the fact that women issue and gun control issue, which she invests heavily in, is not generating many 'likes.'[15]

[14]http://abcnews.go.com/PollingUnit/voted-live-hampshire-primary-exit-poll-analysis/story?id=36805930.

[15]After her defeat in the New Hampshire primary, Clinton started to consider adjusting her strategies. For details, please see http://www.wsj.com/articles/hillary-clinton-cast-in-underdog-role-as-new-hampshire-votes-1455041196#livefyre-comment.

Table 5: Tactic Evaluation

| Candidate | Score | Party | Rank |
|---|---|---|---|
| Sanders | 295.46 | D | 1 |
| Trump | 236.63 | R | 2 |
| Rubio | 147.69 | R | 3 |
| Cruz | 117.88 | R | 4 |
| Clinton | 89.81 | D | 5 |

## Conclusions

We have presented a framework to measure, evaluate and rank the campaign effectiveness. We have studied the tactics and tallies of the ongoing 2016 U.S. presidential election. Using Twitter data collected from Sept. 2015 to Jan. 2016, we have uncovered the tweeting tactics of the candidates and second we have evaluated the effectiveness of the candidates' tactics using negative binomial regression and exploiting the variations in 'likes.' We have also applied the Forward-Stepwise selection algorithm to select the most salient topics for the candidates. Lastly, we have calculated the expected number of 'likes' that each tactic is generating and ranked them.

We observe that while Ted Cruz and Marco Rubio put considerable weight on President Obama, their tactics are not well received by their supporters. We also demonstrate that Hillary Clinton's tactic of linking herself to President Obama resonates well with her supporters but it is not so for Bernie Sanders. We also show that Trump is now a major topic for all the other presidential candidates and that the women issue is equally, if not more so, emphasized in Sanders' campaign as in Clinton's.

Our study shows two possible ways that politicians can use the feedback mechanism in social media to improve their campaign: (1) use feedback from social media to improve campaign tactics within social media; (2) prototype policies and test the public response from the social media.

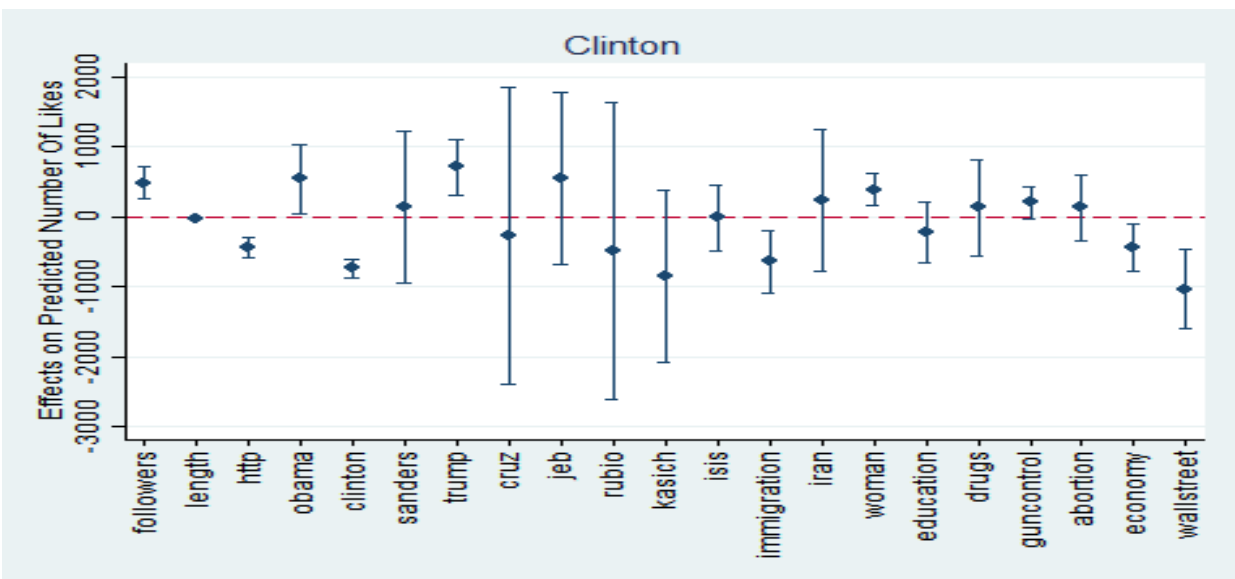

Figure 8: Topic Effects of Clinton's Tweets.

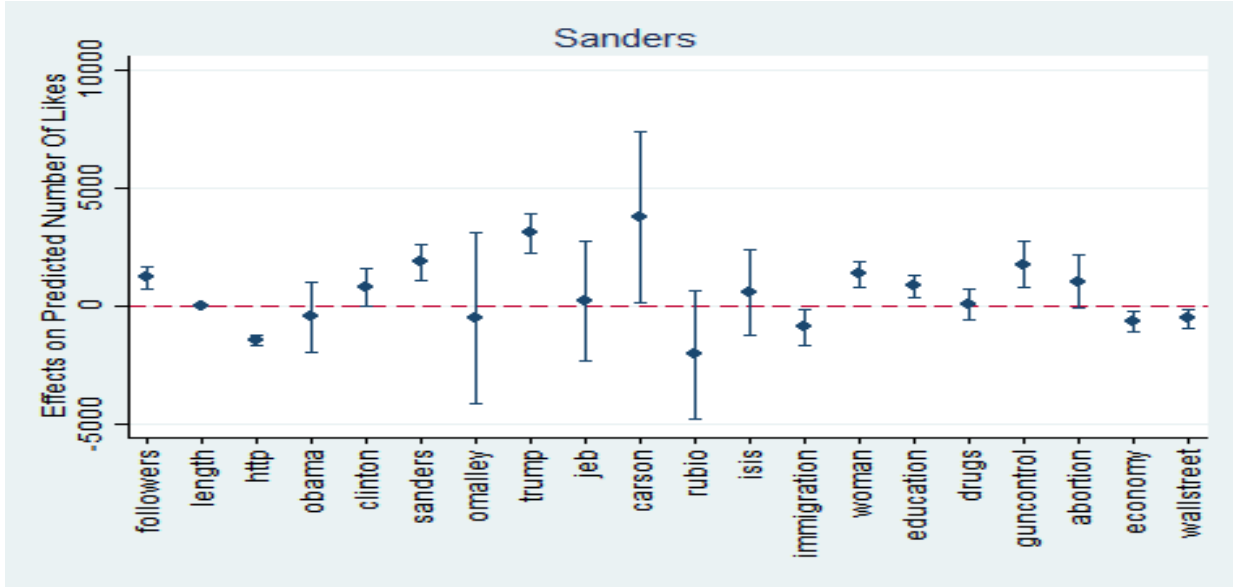

Figure 9: Topic Effects of Sanders's Tweets.

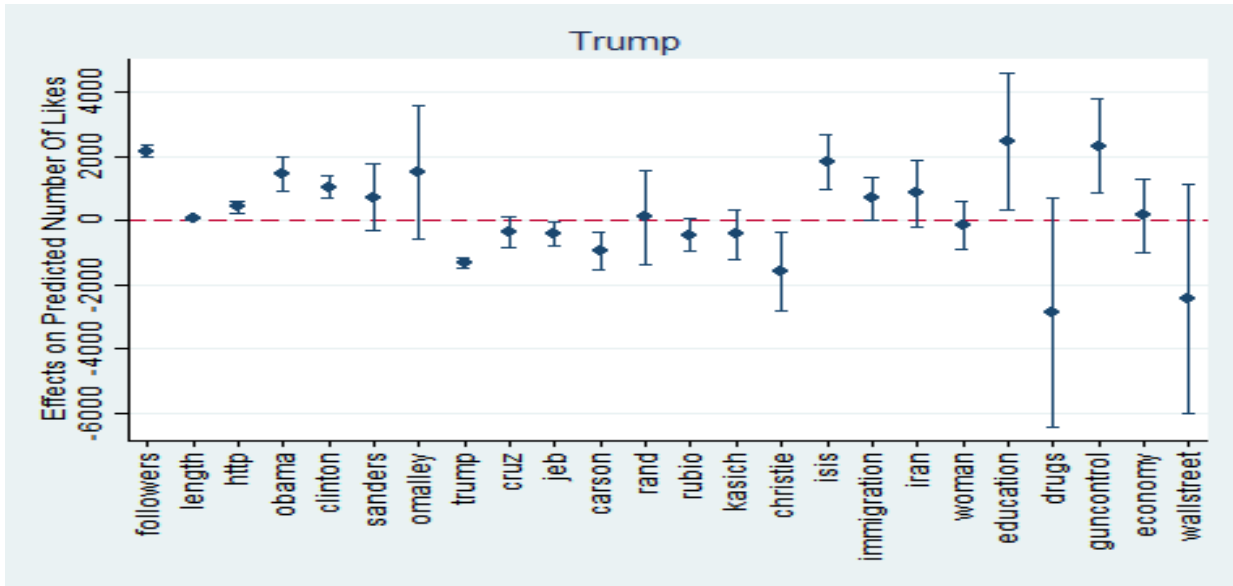

Figure 10: Topic Effects of Trump's Tweets.

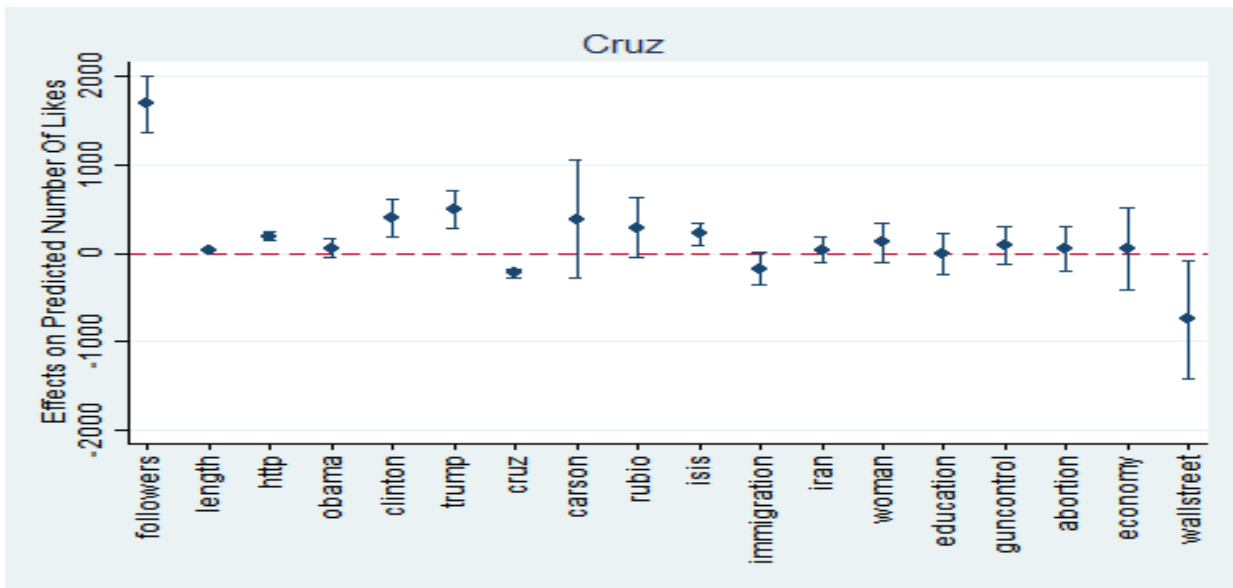

Figure 11: Topic Effects of Cruz's Tweets.

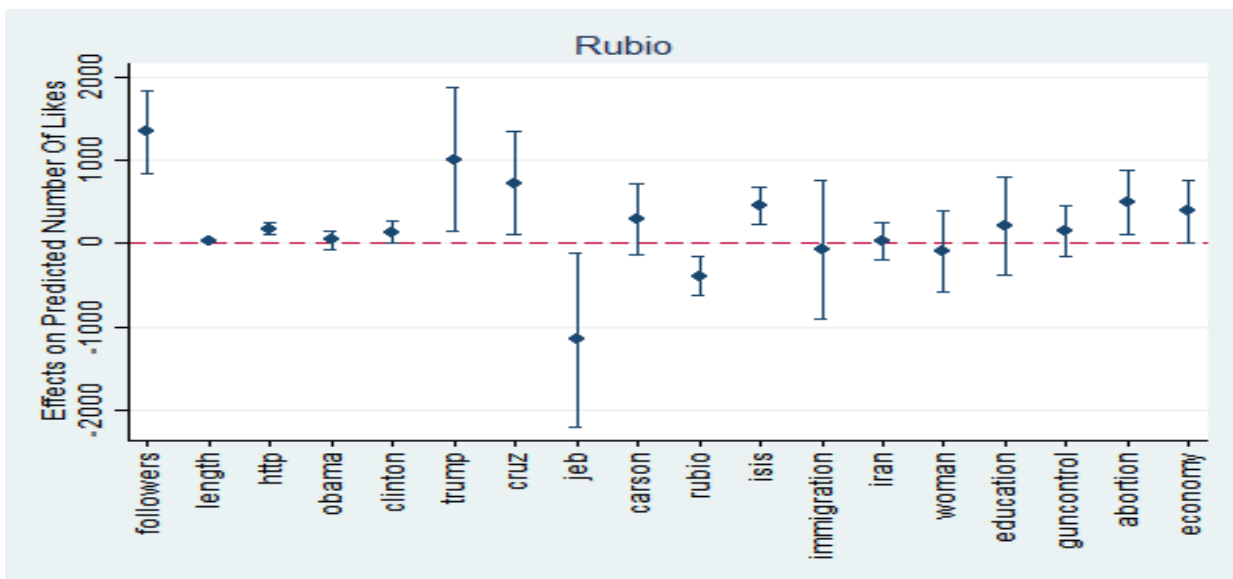

Figure 12: Topic Effects of Rubio's Tweets.